# AIDSS-HR: An Automated Intelligent Decision Support System for Enhancing the Performance of Employees

[1]Nana Yaw Asabere, [2]Nana Kwame Gyamfi

[1]Computer Science Department, Accra Polytechnic, Accra, Ghana

[2]Computer Science Department, Kumasi Polytechnic, Kumasi, Ghana

## Abstract

The performance of employees in an organization is a very important issue for effective delivery and output. Various performance management systems with the aid of Information and Communication Technology (ICT) are currently being used by companies. Such systems are most of the time connected and accessible through to the internet/www. Through review of relevant literature and a system development methodology, this paper proposes an Automated Intelligent Decision Support – Human Resource (AIDSS-HR) system that seeks to control and manage employee activities by tracking the number of years a staff has been at post, keeping inventory on logistics, analyzing appraisal reports of an individual staff and invoking real time prompts devoid of false alarm. The implementation of AIDSS-HR will improve the performance management of employees and benefit the organization, employees and developing nations as a whole.

*Keywords*: *AIDSS-HR, Automated Employee, ICT, Organization, Performance Management.*

## 1. Introduction

Organizations in both developing and developed countries usually have an objective of making sure that the performance of their employees are standard and good enough to meet the right requirements so that expected outputs are produced and/or delivered. In order for organizations to improve the performance of employees, negative factors that affect such procedures should be eliminated manually or by the use of an automated system. Research has revealed that employee complacence and lack of mission goals are some of the setbacks of employee performance in organizations [1][2]. For example in Accra and Kumasi Polytechnics in Ghana, department secretaries and office assistants are transferred every two years from one department to another. This helps such employees use their previous department experience to improve their working performance since long stay of staff at a particular section sometimes brings about reluctance of duty at work. It has also been realized that due to familiarization with work environment and lack of logistics, output of work or productivity becomes poor.

Dissatisfaction of employees at their current place of work also impacts negatively on employee performance [1][2].

Finding solutions to these issues is the ideal case but this does not come that easy due to fatigue and human error. Sometimes there is too much work load on Human Resource (HR) Managers and this leads to lack of policy implementations to improve staff performance. Some HR Managers may also deliberately ignore employees who are not performing well to satisfy their interests. There is therefore the need for the management of various organizations to take into consideration these factors, analyze them and provide a real time Automated Intelligent Decision Support System (AIDSS) to give suggestions as to the possible solutions. Research shows that there are other HR Management Systems that provide various solutions to other challenges in organizations.

The aftermath of such research works led to the proposal of AIDSS in this paper. The main aim of this research paper is to propose a system that can be use to control and manage employees activities, in order to improve their performance in an organization. By taking into account and tracking the number of years a staff has been at post, keeping inventory on logistics, analyzing appraisal reports of an individual staff and invoking real time prompts devoid of false alarm just to name but a few. The proposed AIDSS-HR system will enhance employee performance and produce expected results of organizations in a developing country such as Ghana.

Aguinis [1] defines Performance management as a continuous process of identifying, measuring and developing performance in organizations by linking each individual's performance and objectives to the organization's overall mission and goals. We observe this definition and use it to fulfill our objectives for this paper.

The rest of the paper is organized as follows, Section 2 elaborates on Literature Review, Sections 3 and 4 discuss our Research Objectives and Justification respectively. Section 5 discusses the Research Benefits, Section 6 elaborates on our Proposed System Breakdown Structure (PSBS) and Section 7 elaborates on the Proposed AIDSS-

*HR*. The paper is finally concluded with a recommendation in Section 8.

## 2. Literature Review

This section briefly reviews literature and verifies some existing performance management systems that use ICT. Most ICT systems used for employee performance have been implemented through the web/internet.

According to Kim [3] electronic government is creating the complex challenges of managing an effective Information Technology (IT) workforce in the public sector. Government service delivery is undergoing rapid changes because of IT tools (e.g. Internet and Geographical Information Systems) that are being used at all levels to improve external collaboration, civic engagement, networking and customer service. In reaction to poor performance issues, companies will sometimes offer their employees top-notch training that has little or no effect on the participants' job performance. Management may blame the ineffectiveness of the training in the training program or the trainer, when in fact the training effort was not the correct resolution to the problem in the first place. If training is definitely not the answer, the trainer must identify the root cause (or causes) of the problem and pass this information on to management [1][2]. To achieve "satisfactory" or "exceeds" performance objectives, an employee should have; ability, knowledge, skill, and motivation. Employees should also meets standards, provide feedback and also have a favorable environment. Although all of these factors are crucial to an employee's success on the job, only one aspect which is, knowledge and skill can actually be improved by training. If any of the other factors are the cause of decrease in performance, management or other forces in the organization must institute the changes necessary to resolve the problem [1][2].

Research has revealed that various forms of automated systems have been used in the area of performance management for other uses. Among them are the Contingency Theater Automated Planning System (CTAPS)[1] which was established in 1987 to meet a CAF requirement for a rapidly responsive Command, Control, Communications, Computers and Intelligence (C4I) system to automate and connect elements of the TACS, connect to other organizations or agencies, and permit sharing of common data, and generate, disseminate, and execute tasking orders and coordination messages. The program management directive was to modernize and/or replace existing Air Support Operations Center (ASOC) equipment and to develop a unit level capability. The program has since expanded to accommodate the requirements of all ground elements of the TACS (TACS 1987). Research also shows that effective performance management solutions improve employee goal planning, career development, competency assessment, performance appraisal, compensation management and organizational alignment [1][4]. With effective performance management softwares, organizations can automate performance management to improve employee engagement, retain top performers, and improve performance at both the individual and organizational levels [5]. When organizations align their workforce to key goals and performance measures through performance management softwares, they can identify career paths for employees, create development plans, and identify developmental resources, tasks and ideas to encourage individual development and enhance organizational performance [1][5].

There are some commercial softwares that can be used to help manage these issues. ***Kenexa*** [5] is a company that helps organizations have access to such technology and softwares. ***Kenexa*** offers solutions that enable one's organization to enhance employee engagement, increase productivity, streamline processes, enhance the effectiveness of managers, increase accountability and leverage performance data to make strategic decisions based on a holistic view of their workforce. Kenexa also offers a total Performance Management solution that fully integrates Performance Management and Succession Planning, Compensation Management, Career Development and Goal Management. Through implementation and usage of Performance Management softwares, developed by ***Kenexa*** [5] alignments can be created in organizations, employees are more engaged and organizations can achieve high retention of top performers.

According to [1], an organization that links Performance Management to their recruitment processes, employee assessment and survey action planning can fully leverage employee and organizational data to drive employee engagement and improve overall performance. Other HR systems include OrangeHRM (OHRM)[2] and Carval HR Unity (C-HR)[3], just to name but a few. A performance-recruitment relationship is illustrated in Figure 1.

## 2.1 Decision Software and Decision Support System (DSS)

It is a clear fact from research that a considerable amount of work has been done to reduce or eliminate the poor employee performance in the human resource department of various global organizations. One other problem identified in this area of research is that although existing systems try to manage employees; most of them are currently not intelligent enough to analyze data and provide possible causes and solution to problems. The proposal in this paper therefore finds an improved manner of handling such situations. A better approach will be to deploy an *Automated Intelligent Decision Support System (AIDSS).*

---

[1] http://www.fas.org/man/dod-101/usaf/docs/aoc12af/part10.htm

[2] www.orangehrm.com

[3] http://www.axia-consulting.co.uk/html/payroll_hr.html

Decision Softwares are special kinds of algorithmic softwares designed to help individuals make decisions. Decision Softwares can help users make a decision on a complex problem. Typically, the aim is to choose the best out of several different alternatives. Decision Softwares examine data given to them and, like expert systems, suggest an optimum decision or conclusion. Such decision-making programs are implemented with a number of widely known mathematical decision algorithms [6]. A Decision Support System (DSS) is a collection of Programs/Decision Softwares used for decision-making. Such programs help the management of organizations in planning, forecasting and managing large and complex issues regarding HR. Many DSSs usually include a modeling capability that enables mathematical simulation of a situation to be built in order to test various tactics, intelligence and strategies. Once the model of the system is built by the system developers, various approaches can be tested [6].

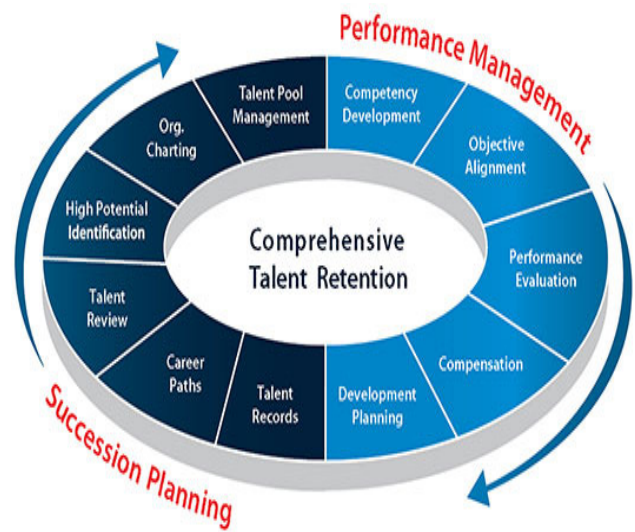

Fig. 1: Performance – Recruitment Management

## 3. Research Objectives

As elaborated in Section 1, the main objective of this research paper is to propose a system that can be use to control and manage employees activities in order to improve their performance in an organization. Due to resource constraints, this research will be based extensively on looking at possible ways of tracking the numbers years a staff has been at post, keeping inventory on logistics, analyzing appraisal reports of individual staff activities and generating real time prompts devoid of false alarm. In order to fulfill the above objectives, our main research questions are as follows:

i    What are the main negative factors that affect employee performance in organizations?
ii   How can ICT be used to develop AIDSS in order to help HR management improve in employee performance?

## 4. Justification of Research

The computer science discipline of Artificial Intelligence (AI) has had tremendous and numerous uses globally. AI has been used in medicine, manufacturing and of late in Management Information Systems (MIS) as well as knowledge based systems with huge success rate stories. An Automated Intelligent Decision Support Systems (AIDSS) is a term that describes a DSS that make extensive use of AI in a more automated, intelligent and improved procedure which can be used to provide solutions to some of the HR management challenges of employee performances in an organization. It is therefore important to note that this research seeks to propose such a solution.

## 5. Benefits of Research

Ghana and other developing countries with a low level rate of ICT penetration will benefit from the use of MIS in the form of *AIDSS-HR* which will help boost the performance of staff in organizations and also increases productivity as well. The recent discovery oil in Ghana will create a magnetic force that will draw all manner of businesses into the country, and with the world moving towards virtual organization paradigm, it will be prudent to have such a proposed system in place to serve its purpose not only to Ghanaian businesses but international businesses as well as multinational companies.

## 6. Proposed System Breakdown Structure (PSBS)

Our Proposed System Breakdown Structure (PSBS) which is made up of the main development activities and the auxiliary activities such as documentation and feasibility studies are listed below.

- Feasibility Studies (FS)
- User Interface Module (UIM)
- Data Input and Editing Module (DIEM)
- Intelligent Analyzer & Auto Alert Module (IAM)

- Documentation

PSBS considers the phases of the software lifecycle and excludes the training and support phases, which are implementation phases. PSBS is represented in a pictorial form shown below in Figure 3.

# 7. Proposed System Architecture

## 7.1 System Requirements

The resources required for our proposed system include both hardware and software comprising of the following:

a) A Computer System - A PC equipped with *AIDSS-HR* should have a Microsoft Windows OS comprising of either Microsoft: Windows Vista, Windows 7 or Windows 8. In terms of Hardware, the requirements for the PC should be: a Liquid Crystal Display (LCD) monitor with a very good resolution, System Unit with an Intel Pentium/Celeron Central Processing Unit (CPU) 2.7 GHz Dual Core, a Random Access Memory (RAM) of 6-8 GB and Hard Disk Capacity at a minimum of 400-500 GB.
b) Software Development Kit (SDK) – JAVA.
c) Database Management System (DBMS) – Oracle or SQL.
d) Test data

## 7.2 System Development Methodology

We propose the Incremental Model which is an evolution of the waterfall model. In this model the product/system is designed, implemented, integrated and tested as a series of incremental sessions. It is a popular software evolution model used in many commercial software companies and system vendors [7]. Incremental software development model may be applicable to projects where:

- Software Requirements are well defined, but realization may be delayed.

- The basic software functionalities are required early.

Figure 2 illustrates the steps that must be followed in the System Development Incremental Model. The advantage of working with this model among others, is that it generates working software quickly and early during the software life cycle. It is also more flexible i.e. it is less costly to change scope and requirements. It is also easier to test and debug during a smaller iteration. Above all it is easier to manage risk because risky pieces are identified and handled during its iteration. However this does not mean that it is a perfect model because it has some setbacks. Each phase of an iteration is rigid and do not overlap each other. Problems may also arise pertaining to system architecture because not all requirements are gathered up front for the entire software life cycle [7].

## 7.3 System Components

The system components pertaining to *AIDSS-HR* are described below:

**User Interface (UI):** The User Interface (UI) is the component that allows interaction between humans and machines. The goal of interaction between a human and a machine at the UI is the effective operation and control of the machine, and feedback from the machine which aids the operator in making operational decisions.

**Intelligent Analyzer (IA):** This is an interface that analyzes all available data on the five main objectives on a daily basis for the HR Manager to act on as an *AIDSS-HR* tool for decision processes. The IA consists of two main components, the knowledge base and rule based components which are used to store the knowledge of employee performance as well as the rules used to check and enhance employee performance.

**Data Input/Edit Module (DIM):** DIM is a component used to receive or accept data and modifies them if there are any errors.

**Data Repository (DR):** DR is a pennant storage location where all refined data are stored.

**Data Protection Module (DPM):** DPM is the component connects *AIDSS-HR* to the web and is also responsible for security of all data stored.

Figure 4 depicts the architecture of *AIDSS-HR*.

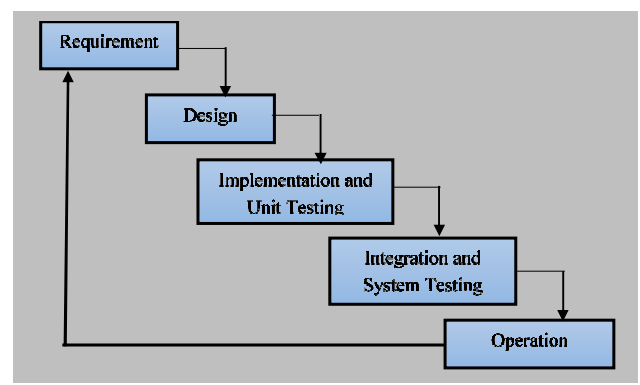

Fig. 2: System Development Incremental Model

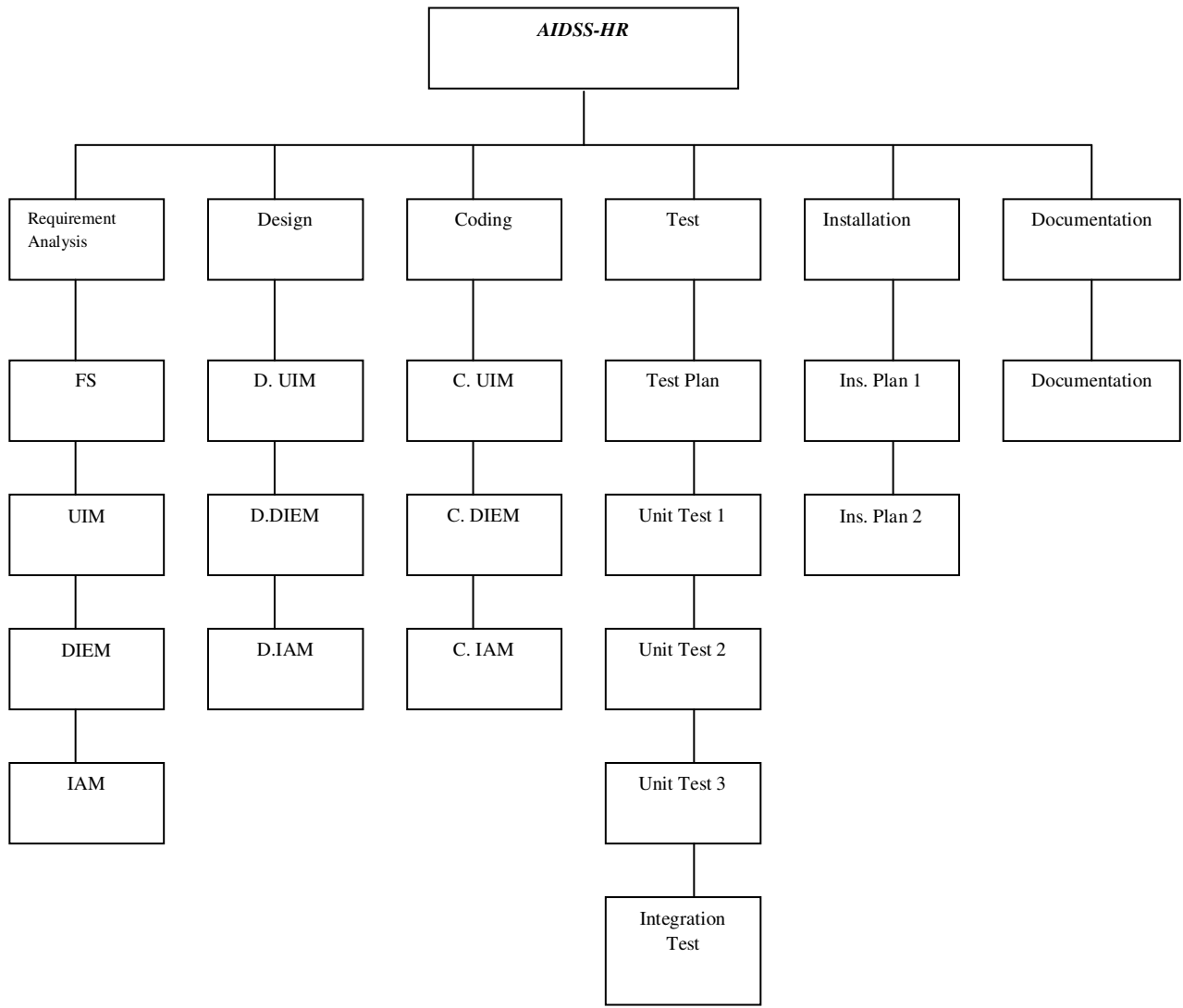

Fig. 3: Proposed System Breakdown Structure (PSBS)

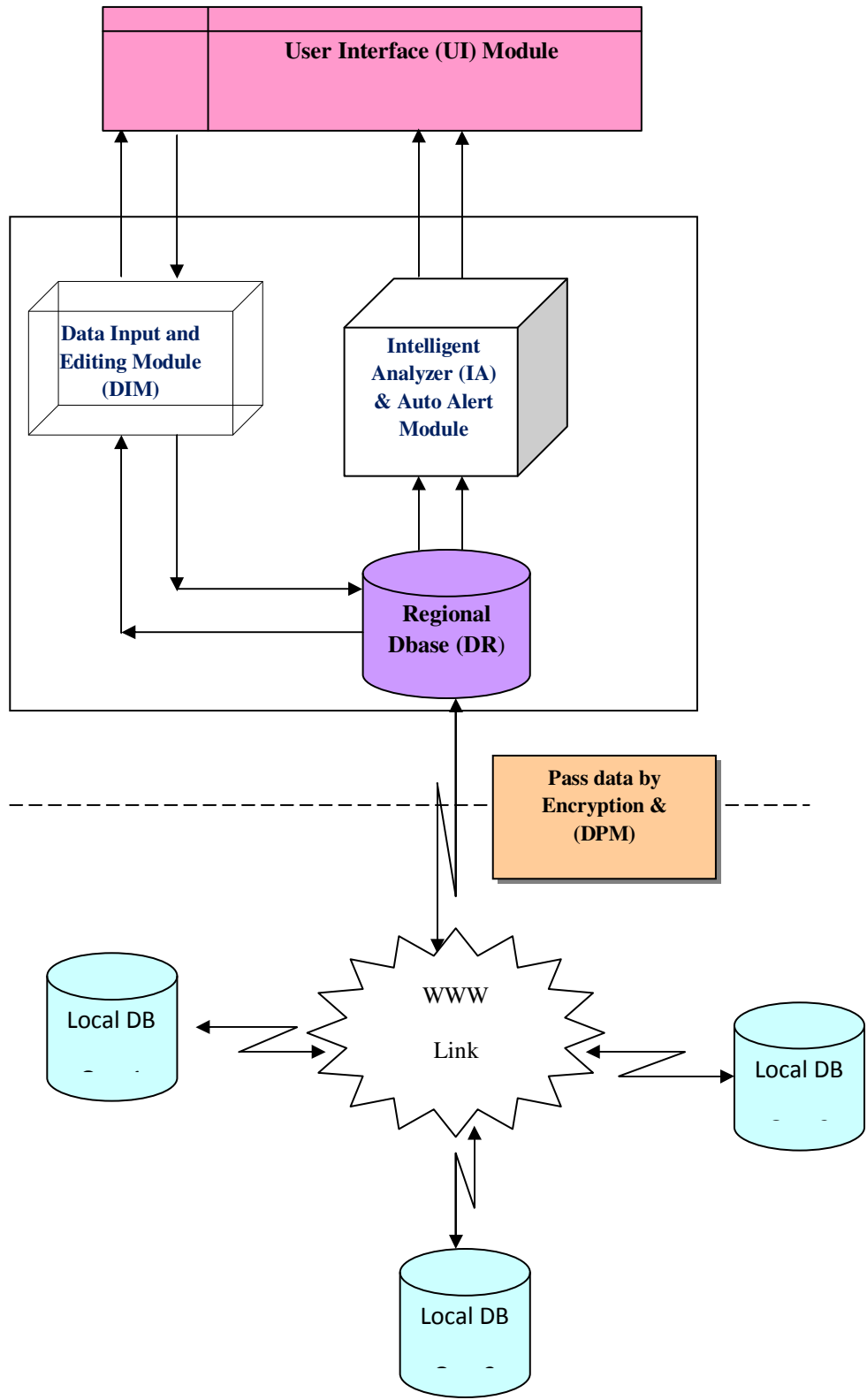

Fig. 4: Proposed Architecture of AIDSS-HR

# 8. Conclusion and Recommendation

The success of every organization worldwide depends on the level of performance management of its employees. Performance management of employees as described in the literature of this paper is a very important issue which HR Managers have to consider daily. Through relevant literature, this paper briefly revealed that most ICT performance management systems are linked to the web/internet and although these systems are performing well to some extent, they are not automated intelligently. This paper therefore proposed *AIDSS-HR*: an Automated Intelligent Decision Support System which can improve some of the flaws of existing systems. Successful implementation of *AIDSS-HR* will improve HR situations such as tracking the number of years a staff has been at post, keeping inventory on logistics, analyzing appraisal reports of an individual staff and invoking real time prompts devoid of false alarm. This paper therefore recommends various organizations in developing countries to establish and implement automated intelligent systems such as *AIDSS-HR* in the future to solve various HR problems in organizations and ease burden on HR Managers.

## Author Profiles

**Nana Yaw Asabere** received his BSc in Computer Science from Kwame Nkrumah University of Science and Technology (KNUST), Kumasi, Ghana in 2004 and MSc in ICT from Aalborg University, Denmark in 2010. He has nine (9) years of teaching/lecturing experience at the tertiary level of education in Ghana and is currently on Lectureship Study Leave granted by Accra Polytechnic, Ghana pursuing his PhD in Computer Science at School of Software, Dalian University of Technology, Dalian, P.R. China.

Nana Yaw has a number of publications to his credits in International Journals and his research interests include: Artificial Intelligence (AI), Software Engineering, Expert Systems, Mobile Learning, E-learning, ICT in Education, ICT for Development, Information Systems, Multimedia, Recommender Systems, Social Computing, Wireless/Data/Mobile Communication and Computing Technologies.

**Nana Kwame Gyamfi** received his BSc in Information Technology (IT) from Presbyterian University College, Abetifi, Ghana in 2008 and MSc in IT from Sikkim Manipal University, India in 2011. He has three (3) years of teaching/lecturing experience at the tertiary level of education in Ghana. Nana Kwame's has been involved in quite a number of IT projects involving software development and IT infrastructural development for the past six years. His research interests include: IT/Network/Computer Security, Computer Network Systems and Management, Mobile Networks and Technologies, DBMS, ICT for Development and Using Web Programming Languages. He is currently a Lecturer, Academic and Industrial Liaison Officer of the Computer Science Department of Kumasi Polytechnic, Ghana.